\documentclass[namedreferences]{solarphysics}
\usepackage[optionalrh,solaromanenum]{spr-sola-addons} 
\usepackage{graphicx}        
\usepackage[usenames,dvips]{color}           
\usepackage{flafter}
\usepackage[colorlinks=true,citecolor=ForestGreen,breaklinks=true]{hyperref}
\usepackage{breakurl}

\usepackage{solaheaderSOLA2138R2}	
\newcommand{\solareceiveddate}{7 November 2012} 
\newcommand{\solaaccepteddate}{8 April 2013}
\newcommand{\solaonlinedate}{10 May 2013}
\newcommand{\Xdoi}{s11207-013-0308-6}

\newcommand{\grl}{    {\it Geophys. Res. Lett.}}

\newcommand{\jgr}{    {\it J. Geophys. Res.}}

\newcommand{\solphys}{{\it Solar Phys.}}

\begin{document}

\begin{article}

\begin{opening}

\title{From Predicting Solar Activity to Forecasting Space Weather: Practical examples of Research-to-Operations and Operations-to-Research Invited Review\\ {\it Solar Physics}}

\author{R.A.~\surname{Steenburgh}$^{1}$\sep
        D.A.~\surname{Biesecker}$^{1}$\sep
        G.H.~\surname{Millward}$^{1, 2}$  
       }
\runningauthor{R.A. Steenburgh \emph{et al}.}
\runningtitle{Practical Examples of R2O and O2R in Space Weather Forecasting}

   \institute{$^{1}$ NOAA/Space Weather Prediction Center, Boulder, Colorado, USA
                     email: \href{mailto:robert.steenburgh@noaa.gov}{robert.steenburgh@noaa.gov} email: \href{mailto:doug.biesecker@noaa.gov}{doug.biesecker@noaa.gov}\\
             $^{2}$ Cooperative Institute for Research in Environmental Sciences (CIRES), University of Colorado, Boulder, Colorado, USA
                     email: \href{mailto:george.millward@noaa.gov}{george.millward@noaa.gov}       
             }

\begin{abstract}
The successful transition of research to operations (R2O) and operations to research (O2R) requires, above all, interaction between the two communities. We explore the role that close interaction and ongoing communication played in the successful fielding of three separate developments:  an observation platform, a numerical model, and a visualization and specification tool.  Additionally, we will examine how these three pieces came together to revolutionize interplanetary coronal mass ejection  (ICME) arrival forecasts.  A discussion of the importance of education and training in ensuring a positive outcome from R2O activity follows.  We describe efforts by the meteorological community to make research results more accessible to forecasters and the applicability of these efforts to the transfer of space-weather research.  We end with a forecaster ``wish list'' for R2O transitions.  Ongoing, two-way communication between the research and operations communities is the thread connecting it all.
\end{abstract}
\end{opening}

\section{Introduction}
     \label{S-Introduction}
     
     Over the past decade, forecast operations at the U.S. National Oceanic and Atmospheric Administration's (NOAA) Space Weather Prediction Center (SWPC) were revolutionized.   Among other factors, the revolution was propelled by the availability of new data, such as those from NASA's \emph{Solar TErrestrial RElations Observatory} (STEREO) \cite{Kaiser2008} and \emph{Solar Dynamics Observatory} (SDO) \cite{Pesnell2012} missions, and from the introduction of physics-based numerical models such as the Wang--Sheely--Arge (WSA)--Enlil model \cite{Odstrcil2004}.  
     
     This revolution was facilitated by successful research-to-operations (R2O) transitions of researcher knowledge and modeling to the forecasting office.  A key component of successful transition is for the space-weather research community to become familiar with operational space-weather forecasting.  Consequently, we will begin by describing the current environment and challenges faced by the practitioners.  Then we will examine three examples of productive space-weather R2O transitions and the interactions between the operations and research personnel (\emph{i.e.} O2R) that contributed to the success.  When these interactions are a two-way street, success becomes much more likely.  A discussion of steps to achieve a successful R2O transition, an example of how SWPC implements R2O/O2R, and the current needs within the operations community, follows.
 
\section{The Practice of Space Weather Forecasting}
	\label{S-forecasting} 
	
   Like their meteorologist counterparts, space-weather forecasters must cope with the rigors of shift work.  They must assimilate several disparate, and often interrupted, data streams as well as a growing number of numerical models.  From these, they are expected to create an accurate description of the current and expected state of the geospace environment.  They are confronted with situations in which they must make significant decisions with incomplete data and inadequate time.  They are simultaneously confronted with the pressure not to miss a warning and the pressure not to trigger a false alarm.  They must be cognizant of a variety of conceptual models that describe space-weather phenomena such as coronal mass ejections and flares.  They must know how and when to apply these models.  If the observations do not fit the models, forecasters must determine why and what impact that will have on the forecast. Then, they must clearly communicate the forecast to a large and varied customer base, keeping in mind the impacts of greatest importance to each customer.  Forecasters must strive to stay abreast of the latest developments in research.  They must continually gain and maintain proficiency in new concepts, tools, and observations introduced to the forecast process.  The professionals who confront these challenges are drawn from a variety of educational and occupational backgrounds.
   
	While the skills and education required to become a scientific researcher are well known (\emph{e.g.} a PhD degree in an associated discipline), less well known is the typical skill set and education of a National Weather Service (NWS) Space Weather Forecaster.  For the two communities to work together successfully, it is paramount that each understand the capabilities and language of the other.  The SWPC forecasters are classified as physical scientists.  This classification requires a four-year undergraduate degree in physical science, engineering or mathematics including 24 semester hours (six to eight university courses) of physical science and/or related engineering science coursework; or an equivalent combination of education and experience.  Currently, the educational background of the forecasters ranges from undergraduate through doctoral level.  Experience levels range from over three decades to less than a year.  Such a diverse group presents significant challenges to researchers attempting to design and deliver education and training.  For instance, the familiarity and level of comfort with mathematics varies among forecasters.  Effective training translates the mathematics into words and concepts without the gory details.  The gory details should, however, be included for those forecasters who are comfortable with them and wish to know more.
   
   The educational background of forecasters can vary significantly, but the demands on space-weather forecasters are nearly identical to those of meteorologists.  So it should come as no surprise that many of the space-weather forecasters have backgrounds in meteorology.  The skills needed to forecast terrestrial weather and provide timely watches, warnings, and alerts are easily transferred to space-weather.  Physicists, environmental scientists, and others round out the current forecast team.  Additionally, the United States Air Force maintains a small contingent of solar observers and space-weather forecasters and some have become forecasters at the SWPC following their military service.
   
\section{Recent R2O, O2R Examples} %
      \label{S-r2o}
	
	Creating operational products and services from the fruits of research has been a renewed focus of the SWPC since it was renamed from the Space Environment Center in 2007.  Recent examples include two ionospheric products, D-RAP \cite{Sauer2008} which provides HF propagation forecasts and US-TEC \cite{Fuller-Rowell2006} which provides total electron content information over the United States and one interplanetary propagation product, WSA--Enlil, which will be covered in detail in later sections.  However, R2O and O2R activities had been ongoing since the SWPC's genesis as the Interservice Radio Propagation Laboratory in the 1940s.
	
	Physics-based numerical models are not the only research to be transitioned to operations.  New models (\emph{both} conceptual and numerical), new observation platforms, and new tools are all candidates for R2O activities.  In the next section, we will examine the convergence of three of these elements:  a new observation platform, a new visualization tool, and a new physics-based numerical model.
    
\subsection{The STEREO Mission} 
  \label{S-stereo}
  
	NASA's \emph{International Sun-Earth Explorer 3} (ISEE-3): \cite{Tsurutani1979} and NASA's \emph{Advanced Composition Explorer} (ACE): \cite{Zwickl1998} are among the first successful examples of enlisting research spacecraft to provide real-time space-weather monitoring to the operational community.  This concept became an integral part of subsequent endeavors such as the ESA/NASA \emph{Solar and Heliospheric Observatory} (SOHO), and NASA's STEREO and SDO missions.  The idea of using STEREO data operationally was present from its inception \cite{St.Cyr2001}.
 
	Imagery, solar-wind, and particle-beacon data from STEREO became available in the forecast office in March, 2007 \cite{Biesecker2008}.  An overview of the various instruments and products produced by STEREO is provided by~\inlinecite{Kaiser2008}.  Data from the \emph{In-situ Measurements of Particles and CME Transients} (IMPACT) and \emph{PLAsma and SupraThermal Ion Composition} (PLASTIC) instruments were displayed in a format identical to similar data from the ACE spacecraft, making it easy for forecasters to begin using the data.  Forecasting the arrival time and potential impact of coronal-hole high-speed streams is one example of the application of the data.  Forecasters also used IMPACT data to identify regions with potential to produce energetic-particle events, then monitored these suspect regions as they entered threatening longitudes.  The S\emph{un Earth Connection Coronal and Heliospheric Investigation} (SECCHI) imagery was used to monitor active regions about to rotate onto the visible solar disk, and to locate the origin of coronal mass ejections.  These predictions and observations made their way into several SWPC products including the \emph{Report of Geophysical and Solar Activity}, a 24 hour summary and three-day forecast of space-weather conditions, and the \emph{Preliminary Report and Forecast of Solar Geophysical Data}, also known as \emph{The Weekly}.
 
	STEREO imagery and data were a welcome addition to the forecast office, but they were not without challenges.  One of the earliest challenges was the changing perspective of the spacecraft as the mission unfolded.  To overcome this, one of the forecasters created a wooden model which included the Sun, Earth, and approximate positions of the STEREO and SOHO spacecraft.  The STEREO spacecraft could be re-positioned on the model as required.  This model helped both forecasters and SWPC visitors interpret the \emph{Extreme UltraViolet Imager} EUVI and \emph{Outer Coronagraph} COR2 imagery.  Recently, one of SWPCs new forecasters has developed code to overlay a Stonyhurst grid and Earth-relative limb position on the STEREO/EUVI imagery, making interpretation even easier.
	
	While the changing perspective was constant, the receipt of imagery and data from STEREO was not. Occasional gaps in both were common.  Forecasters, particularly those with military experience, were used to operating with such limitations and were thankful for any data they could get.  Plans had already been made for some of the data, however, long before forecasters got their first look at it.
	
 	Approximately two years before the STEREO mission was launched, ~\inlinecite{Pizzo2004} described a geometric localization technique to locate and characterize CMEs using STEREO imagery.  \inlinecite{deKoning2009} described the first application of this technique using STEREO beacon-quality coronagraph imagery.  SWPC forecasters who had previously relied on plane-of-sky speed estimates derived from the SOHO/\emph{Large Angle and Spectrometric Coronagraph Experiment} (LASCO) imagery soon began augmenting their estimates using analysis requested from, and provided by, deKoning.  These refined estimates provided forecasters another data point with which to compare forecasts from    of Arrival \cite{Moon2002} and Hakamada--Akasofu--Fry \cite{Fry2001} as well as empirical prognostic methods.  The advent of the STEREO era coincided with the appearance of a time-dependent, three dimensional, magnetohydrodynamic model of the heliosphere called WSA--Enlil.  It would be the convergence of these two technologies that would pave the way for improved geomagnetic forecasts at SWPC.
  
\subsection{The WSA--Enlil Model}
	\label{S-wsa}

	\inlinecite{Pizzo2011} provides a concise description of the model and its initial integration into operations.  The WSA--Enlil model first appeared in the forecast office in 2008.  The WSA portion of the model \cite{Arge2000} provides a semi-empirical characterization of the base solar-wind field from the National Solar Observatory's (NSO) \emph{Global Oscillation Network Group} (GONG) magnetogram measurements accumulated over a solar rotation.  WSA output, in turn, drives the Enlil (\opencite{Odstrcil2004}, \citeyear{Odstrcil2005}) code, which provides the ambient solar-wind outflow. CMEs can then be injected into this ambient flow and the subsequent evolution observed.  Contributions from the entire community, including support for the development of WSA from the Naval Research Laboratory (NRL) and Air Force Research Laboratory (AFRL), and support for WSA--Enlil model development and testing from the National Science Foundation (NSF), Center for Integrated Space Weather Modeling (CISM), and the NASA/NSF Community Coordinated Modeling Center (CCMC), built the foundation for the WSA--Enlil model transition.

	Initially, the model was run on a development platform at NOAA's Environmental Modeling Center and output displayed on a ``test bed'' machine in the corner of the forecast office.  The test bed environment was utilized to firstly address general model robustness and to assess key run statistics (such as the wallclock run time) that would have an impact on future operational use. In addition this system provided an environment to test and develop suitable graphical products, a process that involved scientists and developers with feedback from forecasters.  It also provided an ideal environment for forecasters to become acquainted with the abilities of the model and for initial attempts at validating the output.
	
	Transitioning the model into operations began by constructing a concept of operations (CONOPS). In the case of WSA--Enlil, the CONOPS consisted of running the model on a repeating 2-hourly cycle. In its simplest ``ambient'' mode, the model run would proceed completely automatically on the National Weather Service operational supercomputers using the latest GONG magnetogram as sole input (the GONG input having been pulled automatically by ftp from NSO servers.) On completion, model output was automatically pulled back to SWPC and placed into a database.
	
	This system would be augmented in the event of a potentially geo-effective CME. In ``CME-based'' mode, analysis of the CME by forecasters (described later) would yield key parameters which would then be stored in the database as a ``CME analysis'' via a web-based interface. The parameters from this analysis would then be parsed by automatic scripts running at SWPC into an input file which was then pushed over to the supercomputers in time for the next 2-hourly run cycle. Run scripts operating on the NWS supercomputers would detect the CME input file and WSA--Enlil would run in ``CME based'' mode, with inputs from both GONG and the CME (or CMEs; the system would have the ability to handle multiple CME events.) As before, on completion, model output would be pulled back to SWPC for analysis by forecasters.
	
	Having arrived at a viable CONOPS, the transition process then essentially involves constructing and testing robust scripts to control all aspects of the run process; the networking of inputs and outputs, the running of the model, the creation of graphical products, and the ingesting of final data into the database.
	
	One technical element of implementing a research model on operational supercomputers that is not obvious, requires adapting the model to use a pre-defined directory structure and set of scripts.  The operational supercomputer hosts dozens of models, and it would be impractical from a support and maintenance perspective to allow them to each do things their own way.  Therefore, these are all standardized, enabling more robust real-time support in case of problems arising.  By early 2011, WSA--Enlil processing had moved to operational supercomputers at the National Centers for Environmental Prediction facilities at Gaithersburg, Maryland and Fairmont, West Virginia, for a year-long trial \cite{Pizzo2011}.  It ran four times each day in ``ambient mode''; no CMEs were injected. Instead, the model was used to forecast the arrival and departure of high-speed wind streams associated with recurrent coronal holes.  On-demand CME runs also were provided at this time, with the analysis of the candidate CME and model initiation handled by researchers at the SWPC.

	Once WSA--Enlil became routinely available, both modelers and forecasters began evaluating its performance.  A critique of model performance was added to a daily briefing conducted each morning by the on-duty forecaster and described in more detail in Section~\ref{S-training}.  Because Enlil was dependent on WSA input, forecasters learned to link errors in WSA characterization of the base of the solar-wind with subsequent shortcomings in the overall model output.  (Recognizing when, and why, a model run had gone ``off the rails'' is an essential part of integrating the output into the forecast routine.)  Conversations between researchers and forecasters figured greatly in the development of this ability in the forecast team.  

	As well as evaluating model performance, forecasters also provided feedback on the depiction of model output.  Changes to the depiction based on human factors considerations often happened in less than a week.  This was rewarding to both the forecasters and the modelers.  The forecasters were better able to interpret model output and the modelers knew their work was appreciated, and more importantly, becoming part of the forecast process.  Refinements to the output depiction are ongoing, but the pace has slowed since the initial fielding.
	
\subsection{The CME Analysis Tool (CAT)}
	\label{S-cat}
		
	Since the ultimate aim was to use WSA--Enlil to improve the accuracy of CME arrival forecasts, the focus then became determining how best to inject CMEs into the model output.  \inlinecite{Xie2004} described a cone model for halo CMEs.  A graphical user interface (GUI), called the Cone Tool, was built to allow CMEs to be characterized using this cone model.  Joint validation and verification (V\&V) efforts were subsequently carried out by researchers and forecasters.  Among other things, it became apparent that the relationship between cone angle and radial distance was not constrained in any way, leading to significant variations in estimates of the velocity and angular width for a given CME.
	
	These problems led to the development of a new version of the GUI called the CME Analysis Tool (CAT).  This tool modeled the CME as a three-dimensional lemniscate rather than a cone and used simultaneous comparison of STEREO-A, -B, and SOHO coronagraph imagery to fit the ejecta.  The inclusion of the STEREO data and the choice of the lemniscate had their origins in the work of \inlinecite{Pizzo2004} and \inlinecite{deKoning2009}.  V\&V using the CAT followed, with results more promising than those achieved with the Cone Tool.  The success of CAT suggested the 3D lemniscate model of the CME was a significant improvement over the simple Cone model.  However, since STEREO data will eventually be unavailable for WSA--Enlil input, our eventual goal is to collect enough data using CAT to effectively constrain the cone angle and radial distance obtained with a successor to the Cone Tool. A detailed description of the development of the CAT and the results of V\&V efforts will be presented elsewhere \cite{Millward2013}. Here we will concentrate on the evolution of the CAT through forecaster and researcher interaction.
	
\begin{figure}[h!tbp]
	\centering
	\includegraphics[width=\textwidth]{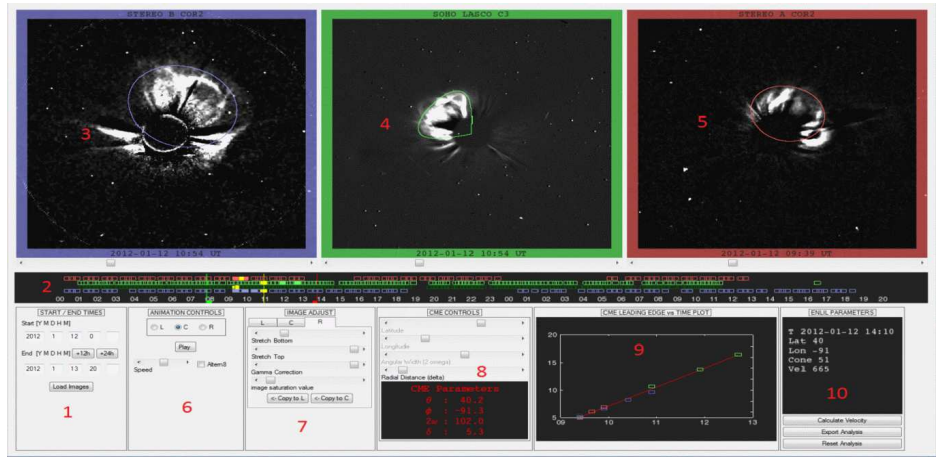}
	\caption{\textbf{The CAT GUI} consist of the (1) Start/End Time Widget; (2) Image Timeline; (3-5) STEREO-B, LASCO, STEREO-A displays; (6) Animation Controls; (7) Image Adjust Widget; (8) CME Controls; (9) CME Leading Edge vs. Time Plot; (10) Enlil Parameters and Export Widget}
	\label{fig:catgui}
	\end{figure}

	The CAT in its present state is shown in Figure~\ref{fig:catgui}.  The CAT GUI consist of the (1) Start/End Time Widget; (2) Image Timeline; (3--5) STEREO-B, LASCO, STEREO-A displays; (6) Animation Controls; (7) Image Adjust Widget; (8) CME Controls; (9) CME Leading Edge vs. Time Plot; (10) Enlil Parameters and Export Widget.  CME analysis and fitting begins with loading the imagery using the Start/End Time Widget.  The Image Timeline is populated with symbols indicating the available imagery from STEREO and LASCO while the images themselves populate the displays.  The analysis step, described below, is often complicated by the delayed receipt of sufficient SOHO or STEREO imagery.  In these cases, a preliminary analysis is conducted with the available imagery and refined as more imagery becomes available.  It is important to note that, with the exception of the NOAA's \emph{Polar} and \emph{Geostationary Operational Environmental Satellite} (POES, GOES) vehicles, all of the space-based platforms upon which space-weather forecasters depend are primarily research platforms.  Consequently, uninterrupted operation and data availability are not guaranteed.
	
	The forecaster can then animate or enhance the imagery as needed to best reveal the CME structure.  Most of the forecaster's time is spent manipulating the CME controls to encompass the ejecta in several frames, that is, adjust the size and orientation of the three dimensional lemniscate to match it to the CME images from SOHO and STEREO, simultaneously.  Once a fit has been accomplished in at least two images at different time steps, the forecaster can produce a leading-edge versus time plot and obtain an estimated velocity. Typically, fits are made to several images.  Once the forecaster is satisfied, the model parameters can be exported.  These parameters are entered into a web-based Solar Predictions Interface (SPI), which prepares the next CME-based model run at the National Centers for Environmental Prediction (NCEP).  In the current concept of operations, the model runs in ambient mode every two hours beginning at 00:00 UTC.  The parameters must be entered before the next even hour, or the forecaster will have to wait until the current ''ambient run'' cycle finishes before the model run can be completed, a maximum of four hours.
	
	Initially, two forecasters were selected to evaluate CAT and attempt V\&V fits for 75 different CMEs.  During the course of the V\&V, the forecasters provided feedback to the researchers regarding the GUI, providing an example of O2R in action.  For example, forecasters soon determined that they needed a way to easily return to default values as they experimented with enhancements to draw out the CMEs in the imagery. Similarly, they needed a functionality that permitted them to quickly discard an unsatisfactory analysis and begin another attempt.  Their suggestions led to the addition of a ``reset'' button on the Image Adjust widget (7) as well as a ``reset analysis'' button on the Enlil Paremeters and Export widget (10).  While not essential to CAT functionality, these enhancements made the tool easier to employ, particularly when the office was busy.  There were also the typical glitches that accompany any R2O effort.  These were captured and reported by the forecasters.  The forecasters were given short-term work-arounds until the code could be remedied. The CAT was eventually installed in the forecast office, promoting experimentation with the tool and practice on CMEs that were not part of the V\&V set.  The parameters that they obtained could then be compared to the researchers results and any systemic problems identified and corrected.
	
	With the advent of routine ambient WSA--Enlil runs on the NCEP computers and the capability to extract model parameters using CAT, real-time CME events were injected and output retrieved and displayed using SPI.  At first, researchers conducted the analysis and prepared CME-based WSA--Enlil runs.  They were on call around the clock and would be notified by forecasters when candidate CMEs were observed.  Forecasters eventually gained more experience and began initiating model runs independently.  Researchers would also initiate independent runs using the characteristics gleaned from their own CAT results.  The forecasters and researchers discussed the confidence in their respective analyses.  Once the WSA--Enlil run was complete, the output was scrutinized, and the solution deemed most probable became the official forecast.  The multiple runs effectively functioned as a mini-ensemble; a valuable contribution to the overall forecast.  Post-mortems were subsequently conducted during which forecasters and researchers discussed the results.  Topics might include strategies for dealing with missing or poor quality imagery; determining what features are, or are not, part of the CME; and coping with overlapping events.  Exchanges helped forecasters gain confidence in the tool and their ability to use it.  Similarly, researchers were able to learn how forecasters were using the tool and correct any misunderstandings or errors.
	
\begin{figure}[htbp]
	\smallskip
	\centering
	\includegraphics[width=\textwidth]{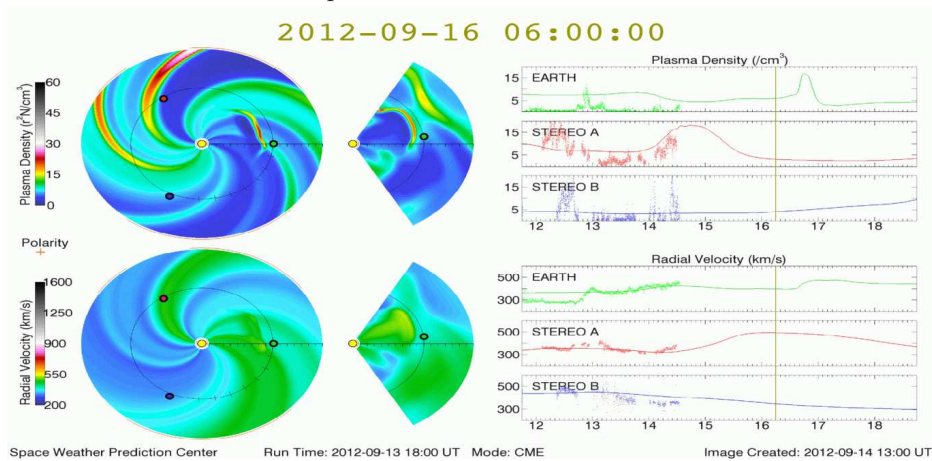}
	\caption{An example of ACE data superposed on WSA--Enlil model output.  The date is depicted on the x-axis.  Earth is depicted on the right side of the plan--view plots and a circle is drawn at the 1AU range.  STEREO A and B spacecraft are also depicted.  In this example, WSA--Enlil solar-wind speed predictions were performing well at STEREO-A and the ACE spacecraft based on the observed data.}
	\label{fig:ace_overlay}
	\end{figure}

\begin{figure}[p]
	\centering
    \includegraphics[scale=.3,angle=90]{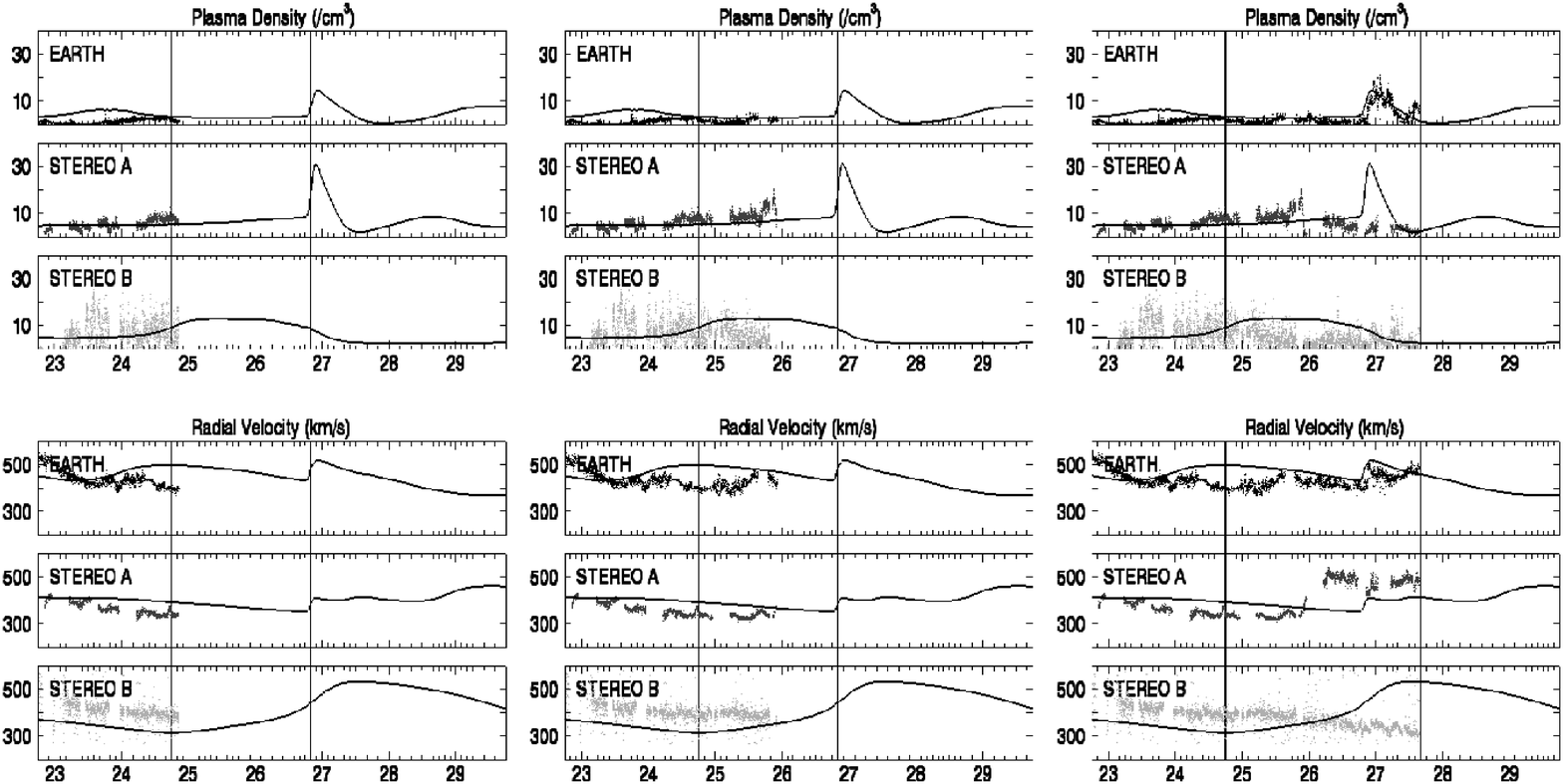}
	\caption{An example of ACE data superposed on WSA--Enlil model output on 24 (left), 25 (middle), and 27 (right) February 2012.  The date is indicated on the x-axis.  The forecast for arrival at Earth verified well; however the ICME arrived at STEREO-Ahead sooner than anticipated.}
	\label{fig:montage}
	\end{figure}

	The model run(s) were typically available one to four days before the CME was predicted to arrive.  Superposing observed ACE solar-wind data and the model prediction during the transit allowed forecasters to determine how well WSA--Enlil was performing and adjust the forecast accordingly.  Figure~\ref{fig:ace_overlay} shows an example in which WSA--Enlil solar-wind speed predictions were performing well at STEREO-A and the ACE spacecraft.  The ability to easily and rapidly compare model predictions with observed values is a key element in crafting an accurate forecast.  Figure~\ref{fig:montage} shows a similar overlay over three days (24, 25, and 27 February 2012).  The predicted speed was higher than the observed speed until late on the 25th, when the observed speed began to match the model forecast.  The ICME arrived at Earth as predicted by the model, but arrived at STEREO-A earlier than expected.

By developing CAT to incorporate STEREO imagery in the specification of CME parameters for WSA--Enlil, we effectively constrained the relationship between the cone angle and radial distance .  Anecdotal evidence suggests that model output has improved as a result and data are being collected and analyzed for a more formal analysis to be presented elsewhere \cite{Millward2013}.

\section{Keys to Successful R2O}

From the field of meteorology, \inlinecite{Doswell1986} suggested several actions for the successful transition of science and technology into operations.  Among them, he recommended having ``researchers work together with forecasters in developing new science and technology to \emph{suit the needs of the forecasters} [emphasis added]''.  This recommendation captures the essence of the entire transition of STEREO, WSA--Enlil, and the CAT to operations at SWPC.  

\subsection{Lessons Learned}
The importance of communication was echoed almost a quarter century later by \inlinecite{Eduardo2009} and comes at the end of his list of ``building blocks'' for successful R2O:

\begin{quote}
	\begin{enumerate}
	\item Identifying the customer's wish list [\emph{i.e.} avoid a model in search of a customer]\label{it:wish} %
	\item Conducting Verification and Validation \label{it:vandv}
	\item Specifying Failures and Events \label{it:fail}
	\item Documenting Errors and Uncertainties \label{it:error}
	\item Flagging -- identifying compromised quality \label{it:flag}
	\item Providing thorough and understandable documentation \label{it:doc}
	\item Communicating -- interaction between researchers and forecasters \label{it:communicating}
	\end{enumerate}
\end{quote}

While this article has largely focused on the interactions between researchers and forecasters, it is important to note that the foundation for success was built on the preceding blocks.  SWPC works closely with the operational user community to anticipate their needs and leverage the work of the research community to meet those needs. The development and implementation of WSA--Enlil grew out of SWPC's desire to move from specification to prediction, and to provide better forecast products for our customers (Item~\ref{it:wish}).  

Items~\ref{it:vandv}--\ref{it:error} led to the abandonment of the Cone Tool and the development of the CAT as described in Section~\ref{S-cat}.  WSA--Enlil model V\&V has been complicated by the variability of forecasters; given the same CME, each forecaster will produce a slightly (or sometimes wildly) different analysis and resulting model parameters.  One goal of training and V\&V was to move the analysis solutions towards a reasonable ensemble and eliminate the outliers.  Since resource limitations meant model runs could not be completed for every set of parameters generated, it was impossible to know which analysis was ``right''.  When multiple runs were accomplished for an operational event, as described in Section~\ref{S-cat}, a less than satisfactory analysis may still result in a better forecast because the ambient solar-wind was not being handled well by the model, i.e. the forecast was right for the wrong reasons.  While WSA--Enlil model output is not flagged (Item~\ref{it:flag}), the overlay of observed data from the ACE spacecraft described in Section~\ref{S-cat} above alerts the forecaster when model output is in disagreement with reality.  Collaboration between the forecast and research staff and resulted in an accurate and usable guide to the CAT (Item~\ref{it:doc}).

There is a temptation to call model output ``the forecast''.  However, it is important to remember that model output is only one ingredient in the forecast.  The forecast is built from observations and analysis, empirical tools, and expertise.  When a forecaster succumbs to the temptation to call model output the ``forecast'', it leads to an abdication of responsibility and erosion of expertise.  This phenomenon was described by \inlinecite{Snellman1977} as ``meteorological cancer'' after numerical weather prediction (NWP) models became operational.  This topic will be revisited in Section~\ref{S-wish}.

A final lesson learned worthy of its own section is the importance of a robust training program to R2O success.  Such a program will enable forecasters to take full advantage of any new observational platforms, tools, or numerical models.

\subsection{The Importance of Education and Training}
	\label{S-training}
	
	Training forecasters with a diverse educational background who work a rotating shift schedule is a non-trivial, but absolutely essential, undertaking.  Throughout the development of WSA--Enlil, the Cone Tool, the CAT, and their integration into forecast--office operations, formal training was conducted with the SWPC forecasters.  Knowledge-based education and training included a review of the quiet corona, structures in the solar-wind, near--Sun magnetic fields, interplanetary propagation, and ``how derived cone parameters relate to WSA--Enlil physical inputs and the subsequent interplanetary evolution.''  Once the relevant background had been established, the focus shifted to learning to use the CAT.
	
	A combination of task-based and knowledge-based training was required for the CAT.  Two training sessions were conducted by the researchers.  During the initial session, they provided background information about the shortcomings of the Cone Tool and the application of the geometric localization technique.  Forecasters were then given the opportunity to fit some example CMEs.  As they practiced, the researchers would guide the forecasters through the process and correct any errors or misunderstandings along the way.  The two forecasters chosen for the CAT V\&V described in Section~\ref{S-cat} were also present to assist. One of them wrote a concise CAT users guide that was available during and after the training.  Because the guide was written by a forecaster for forecasters, it was focused on the operational application of the tool rather than the theoretical underpinnings.
	
	After this initial session, the trainees were provided a list of ten candidate CMEs to fit.  Their solutions were stored and evaluated by the researchers.  A second training session was conducted during which the various solutions were discussed, the outliers identified and any problems corrected, and questions answered.  
	
	In addition to the in-house training, forecasters have attended the Center for Integrated Space Weather Modeling (CISM) Summer School \cite{Lopez2008} since its inception, typically within their first year at SWPC.  This course provides an up-to-date overview of both the theoretical and applied aspects of space-weather.  Forecasters are introduced to space-weather modeling.  Their classmates are primarily future researchers.  Thus, construction of the bridge between research and operations begins early in the careers of both populations.
	
	Since forecasters and researchers are located in the same facility, daily interactions provide additional, mostly informal training opportunities.  One formal interaction is the daily space-weather briefing mentioned in Section~\ref{S-wsa}.  Every weekday morning, the forecaster provides a review of activity over the past 24--72 hours, and a rationale for the day's forecast.  This allows researchers to see what tools, models, and data a forecaster is focused on as well as to judge the rationale being offered.  They also offer insight into what the latest research might say about the situation.  The forecaster provides an evaluation of tools, model performance, and data quality while receiving immediate feedback on intricacies of the same.
	
	Finally, \inlinecite{Stuart2007} note that forecasters and researchers approach learning from two distinct perspectives.  They described researchers as ``knowledge seekers'', suggesting they ``prefer to understand theoretical concepts.'' Forecasters were described as ``goal seekers'', suggesting they ``prefer concrete examples from the real world.''  While it is dangerous to paint entire communities with such a broad brush, including real--world examples in education and training efforts facilitates understanding and application among forecasters.  Practicing on real CMEs during the CAT training described above, for instance, enabled forecasters to immediately grasp the value brought by the new tool.

	Twenty-six years ago, \inlinecite{Doswell1986} wrote of the importance of training to an improved meteorological forecast system:
	
	\begin{quote}
	It is inconceivable that we spend billions of dollars on hardware and virtually nothing on meaningful training.  Technology is supposed to enable us to do as well or better than our present performance without \emph{any} substantial investment in our people.  I think that any real \emph{improvements} in the performance of our forecasting system must come from advances in the concepts and tools provided to the \emph{people} operating the system.  Giving a word processor to someone whose knowledge of English is deficient does not make the writing any better than when it was done with pencil and paper.  And the word processor cannot do the writing by itself.
	\end{quote}

\subsection{Making Research Accessible}
	\label{S-Accessible}
	
As noted in Section~\ref{S-r2o}, numerical models are not the only candidates for R2O activities.   \inlinecite{Siscoe2007} compared the evolution of meteorology and space-weather forecasting, noting the most significant gains in forecast accuracy came with the introduction of numerical weather prediction into operations.  These advancements in model sophistication would not have been possible, however, without an ever-expanding network of observations and robust data-assimilation schemes with which to initialize the models.  Nor would the advancements have been possible without sustained research directed towards understanding the evolution of synoptic and mesoscale phenomena.

	The responsibility for making research accessible rests in both the research and the operations community.  \inlinecite{Doswell1981} described the challenges faced by weather forecasters in this regard:
	
	\begin{quote}
	Forecasters are not generally trained or encouraged to read the current meteorological literature, so typically they can't even try to ferret out those articles that actually have a bearing on their job.  Rotating shift work makes it difficult for forecasters to participate actively in any but the most brief of training programs, much less maintain currency in their profession.  (Having experienced the rigors of shift work, we can only admire those rare individuals who can conduct applied research and/or stay up-to-date under the handicap of rotating shifts.)
	\end{quote}

	In the three decades since that article was written, the NWS has taken steps to improve the situation.  One step was the creation of the Science and Operations Officer (SOO) position at forecast offices and national centers. Among their many duties, ``The [SOO] is expected to:
	
	\begin{enumerate}
	
		\item Initiate and oversee the transfer of new technologies from the research community to the operational environment\ldots
		
		\item Leads and/or participates in significant joint research projects and developmental efforts conducted in a collaborative manner with \ldots science experts in the collocated or nearby university, other Federal agencies, and/or related professional societies and organizations.

		\item Assesses continuing and future training needs required to successfully incorporate new technology and science into the \ldots operations.

		\item Coordinates and consults with scientists in the NWS, NOAA, other agencies, academia and the private sector to identify development opportunities for enhanced forecast procedures and techniques to be used at the [forecast office]. Integrates new scientific/technological advances and techniques into\ldots operational procedures and operations. \cite{Opmsoo2012}''	

	\end{enumerate}
	
	The SOO is only required to work rotating shifts $25\,\%$ of the time, opening up the possibility that they will be able to focus on the primary duties described above.  Space-weather forecast centers would do well to emulate this model.
	
	In addition to the creation of an SOO position, another effort within the meteorological community to bring research to operators is the establishment of the Cooperation in Meteorological Education and Training (COMET) program \cite{Spangler1994,Johnson1996} under the auspices of the University Corporation for Atmospheric Research and NWS.  COMET provides a variety of internet-based and residence courses designed to bring the fruits of research to the operational community.  The on-demand internet-based courses allow forecasters to complete training that they might otherwise miss because of rotating shift work.  The COMET program originally began with a focus on mesoscale meteorology.  However, in the past few years, SWPC has collaborated with COMET developers to create and offer space-weather tutorials on the COMET website.  These tutorials were designed to provide meteorologists with enough background in space-weather to field routine questions from the public.
	
	Finally, traditional methods of making research available such as conferences and publications should not be overlooked.  Assuming at least one forecaster is able to attend a given conference, the possibility exists for mutually beneficial interactions between representatives from the research and operational communities.
	
\subsection{An Operator's Wish List}
	\label{S-wish}

	A discussion of R2O and O2R would not be complete without the inclusion of a wish list from the operations' side of the house.  This short wish list incorporates suggestions from operations personnel both inside and outside the forecast office.
	
	\begin{itemize}
		\item \inlinecite{Siscoe2007} called the stage of meteorology that followed the advent of numerical weather prediction the storm-tracking stage.  The capability to track ICMEs across the heliosphere and, in particular, interrogate the magnetic structure throughout its transit, is a key to evolving to this stage.  Although we are improving the arrival-time predictions through numerical models, we are no closer to being able to forecast the magnitude of the ensuing geomagnetic storm after the ICME reaches Earth.  This is a huge gap.
		\item As numerical weather prediction advanced, grid spacing became smaller and mesoscale models emerged.  Similarly, the creation of geospace models, and subsequently, \emph{regional}, geospace models would be a welcome development.  Several large-scale models have emerged over the past decade and are being evaluated \cite{Pulkkinen2011}.
		\item Current flare probabilities are based on empirical studies.  Being able to accurately forecast regions capable of producing M and X-class solar flares, CMEs, and energetic particle events well in advance, would be a great step forward (see, \emph{e.g.}, \opencite{Falconer2003}).
		\item Closely related is the ability to forecast regions of emerging flux (see, \emph{e.g.}, \opencite{Ilonidis2012}).
		\item The ability to nowcast/forecast the radiation environment in interplanetary space and at GEO, MEO, and HEO.
		\item The creation of tools to help \emph{quantitatively} analyze and track active-region creation, evolution, and decay using space-based imagery.  This includes areal extent, McIntosh classification, Mt. Wilson classification, gradients, and any other characteristics that may lead to forecast techniques.
		\item Finally, the ability to run displaced-real-time simulations of historic space-weather events in an environment that mirrors the forecast office.  This will enable forecasters to maintain proficiency and prepare for the next solar maximum during solar minimum.  \inlinecite{Magsig2003} describes such a system being used to train terrestrial weather forecasters.
	\end{itemize}
	
	 Any technology or automation introduced to the forecast office should support the continued development of expertise, not hinder it.  It should make the forecast process and the resulting forecast better, not worse.  This seems obvious but, unfortunately, has not always been the case in the field of meteorology (\opencite{Pliske1997} and \opencite{Klein2011}).  \inlinecite{Klein2011} describes the situation:
		\begin{quote}
		The NWP [numerical weather prediction] system was so awkward to use that forecasters just accepted what the system said unless it really made a blunder.  The experts forced to use it found themselves accepting forecasts that were sort of good enough.  They know they could improve on the system, but they didn't have the time to enter the adjustments \dots This problem will probably disappear once they retire.  The newest forecasters won't know any better, and so expertise will be lost to the community because systems like NWP get in the way of building expertise.
		\end{quote}
		
	Wrapped up in each of these ``wishes'' is an increased understanding of the phenomena we are trying to forecast.

\section{Conclusion} 
      \label{S-Conclusion}
      
    New observational capabilities (STEREO), methods to exploit these capabilities (Geometric Localization \& CAT), and a robust magnetohydrodynamic model (WSA--Enlil) were integrated to form a powerful new forecast tool.  Throughout the creation of this new tool, dialogue between the researchers developing it and the forecasters who would wield it contributed greatly to its smooth integration into operations.  Some lessons for successful R2O / O2R activities can be drawn from this endeavor.
    
    We have presented three examples of the successful transfer of research to operations.  STEREO, WSA--Enlil, and CAT can each stand alone as a case study in R2O / O2R interaction.  However, by integrating these technologies into a new operational concept, we hope to be able to continuously improve the forecast process and our resulting products.  \inlinecite{Doswell1981} point out that the great contributions in meteorology came ``at those times and in those places [where] the interactions and mutual respect between theoreticians and forecasters was substantial.''  This climate is being actively cultivated at SWPC, leading to positive outcomes within the organization and extending to the larger community we serve.
  
	 In another encouraging development, \inlinecite{Lanzerotti2011} noted the establishment of a ``Research to Operations / Operations to Research'' working group for the recent Decadal Survey \cite{Decadal2012} signaled the recognition of the importance of including the O2R requirements from ``designers, systems operators, modelers, [and] forecasters'' in the ``thinking and discussions by the decadal survey steering committee for new solar-terrestrial research programs, initiatives, and activities.''  This acknowledgement contributes to community cohesion and sets the stage for continued contributions to science and society.

%

\begin{acks}
 The authors thank Vic Pizzo, Eduardo Araujo-Pradere, Howard Singer, and Chris Balch for their helpful contributions.
\end{acks}



%
%

\end{article} 

\end{document}